\documentclass[a4paper,11pt]{article}
\usepackage[utf8]{inputenc}
\usepackage{mathtools}
\usepackage{natbib}
\usepackage{graphicx}
\usepackage[font=footnotesize]{caption,subcaption} 
\usepackage{array} 
\usepackage{listings} 
\usepackage{lineno} 
\usepackage[nottoc]{tocbibind} 
\usepackage{hyperref} 
\usepackage[percent]{overpic} 
\usepackage{float} 
\usepackage{chngcntr} 
\usepackage{color}
\usepackage{authblk}
\usepackage{fullpage}
\usepackage{amsmath,amssymb,dsfont}
\usepackage{amsthm}
\usepackage{placeins}

\renewcommand*{\thefootnote}{\fnsymbol{footnote}}

\begin{document}

\title{Data-Driven Economic Agent-Based Models$^{\star}$}
\author{Marco Pangallo and R. Maria del Rio-Chanona}
\date{\today}

\maketitle
\footnotetext[1]{We would like to thank the co-organizers and participants of the workshop ``Data-driven economic agent-based models'' (\url{https://sites.google.com/view/wddeabm}), where we discussed many of the ideas presented in this chapter.}

\renewcommand*{\thefootnote}{\arabic{footnote}}

\begin{abstract}

Economic agent-based models (ABMs) are becoming more and more data-driven, establishing themselves as increasingly valuable tools for economic research and policymaking.  We propose to classify the extent to which an ABM is data-driven based on whether agent-level quantities are initialized from real-world micro-data and whether the ABM's dynamics track empirical time series. This paper discusses how  making ABMs data-driven helps overcome limitations of traditional ABMs and makes ABMs a stronger alternative to equilibrium models. We review state-of-the-art methods in parameter calibration, initialization, and data assimilation, and then present successful applications that have generated new scientific knowledge and informed policy decisions. This paper serves as a manifesto for data-driven ABMs, introducing a definition and classification and outlining the state of the field, and as a guide for those new to the field. 
\end{abstract}


\section{Introduction}

Economic modeling is essential to understand how the economy works and to address pressing global challenges. While econometric studies provide an initial understanding of causal effects, theoretical economic models enable policymakers to simulate scenarios like tax reforms, industrial policy, or economic crises and their impacts on employment, inflation, GDP, and inequality. Theoretical models serve two additional purposes: first, they help us anticipate how agents might respond in new scenarios like climate change or the rapid adoption of artificial intelligence, and second, to project the effects of policies that could reshape agent behaviors in ways not observed historically \citep{muth1961rational}. 

Traditionally, economic models have relied on the concept of equilibrium, where the state of the economy is derived from solving a set of equations typically based on agents’ maximizing behavior. This approach can model how agents respond to novel situation that may alter their behavior, as agents re-maximize in the new setting. Since equilibrium models are based on equations, they can often be solved analytically, making them clear and transparent. Furthermore, they provide a consistent and unifying framework that reduces ad-hoc modeling assumptions. However, these benefits come at a cost: equilibrium models rely on simplifications that can misrepresent real-world dynamics, heterogeneity, and human behavior. For example, the use of a representative agent limits the ability to capture diverse behaviors, while assumptions of perfect rationality can overlook heuristics common in decision-making. As a result, even though policy counterfactuals in equilibrium models are internally consistent, they may miss critical real-world phenomena, such as sudden crises, inequality, and bounded rationality, which limits their reliability for policy analysis.

In the last 30 years, Agent-Based Models (ABMs) have emerged as an alternative to equilibrium models \citep{axtell2024agent}. ABMs are non-equilibrium in the sense that all model variables are obtained as a map from previously computed values, rather than by solving equilibrium equations \citep{pangallo2024equations}. For instance, cash available to an agent may update following sales and purchases, but neither of these is determined by setting demand equal to supply. Non-equilibrium dynamics makes it possible to include all sort of realistic aspects. For example, an ABM can simulate decision-making through heuristics \citep{artinger2022satisficing}---such as households adjusting spending based on recent income changes rather than from market clearing conditions. As another example, ABMs can easily capture real-world dynamics and transient states, which is crucial for modeling sudden events like the COVID-19 pandemic. However, this flexibility comes with trade-offs: without equilibrium assumptions, ABMs may lose mathematical tractability and often lack closed-form solutions. The flexibility in modeling choices can also make ABM rules appear ad hoc or arbitrary, especially in earlier applications where ABMs served as thought experiments rather than calibrated models of actual economies, limiting their reliability for policy recommendations.

In this chapter, we argue that the recent trend toward data-driven ABMs is helping overcome their traditional limitations, making them a compelling alternative to equilibrium models. First, agent-level micro-data enable us to observe behaviors directly, reducing the number of arbitrary assumptions. For example, in an economic-epidemic model simulating economic and mobility decisions \citep{pangallo2023unequal}, mobility data allow us to avoid detailed assumptions about daily activities like bringing kids to school and commuting, which are not crucial for an economic model. Second, ABMs can now faithfully represent a given economy by building synthetic populations that accurately reflect individual and household attributes, or by reconstructing the production networks linking firms. This detailed representation improves the quantitative reliability of policy counterfactuals, offering more grounded results than studies in abstract economies. Third, ABMs can now track empirical time series data, allowing for in-sample calibration and out-of-sample testing. In some cases, data-driven ABMs have been able to forecast macroeconomic variables, even after major shocks like the COVID-19 pandemic \citep{poledna2023economic,pichler2022forecasting}. This focus on time series fitting and forecasting increases the scientific validity of ABMs by allowing empirical testing of model assumptions. Assumptions that do not improve model performance can be streamlined or removed, simplifying models and making them more transparent.

The tendency to more closely link ABMs and data has been greatly accelerating in the last ten years. In the following, we first discuss in what precise sense economic ABMs are becoming more and more data-driven, and then review new methods and applications, including success stories of recent data-driven ABMs, and highlighting challenges and opportunities for future research.

\section{Definitions}

\subsection{A technical definition of ABMs}
Although there exist several conceptual definitions of ABM \citep{jennings2000agent,bonabeau2002agent,de2014agent,delligatti2018agent}, to assess what makes an ABM data-driven we need a \textit{technical} definition. Here is our proposal.

An ABM is composed of $N \gg 1$ model units, indexed by $i = 1, \dots, N$, that for simplicity we call \textit{agents}.\footnote{Model units could be agents, i.e. autonomous units with ``agency'', or features of the environment.} It is important that they are distinct discrete units, not a ``continuum'' of identical agents.  Each agent is characterized by a vector of time-dependent \textit{variables}, denoted by $\mathbf{x}_{i,t}$ at time $t$, and a vector of fixed \textit{attributes} represented by $\mathbf{a}_{i}$. At the system level, we define a vector of \textit{model-wide} variables $\mathbf{y}_{t}$, which evolve over time, and a vector of constant \textit{parameters} $\mathbf{\theta}$ that govern the behavior of the entire system. An ABM implicitly defines a probability distribution $\mathbb{P}$ that determines the current values of both the set of agent-specific vectors $\{{\mathbf{x}_{i,t}}\}_{i=1}^{N}$ and the model-wide vector $\mathbf{y}_t$, given the past values of these vectors, the agents' attributes, and the parameters, as follows\footnote{While $N \gg 1$ requires that $\{{\mathbf{x}_{i,t}}\}_{i=1}^{N}$ is non-empty, other components in Eq. \eqref{eq:abm} can be empty sets. For example, the model may not include model-wide variables, making $\{\mathbf{y}_{\tau<t}\}$ an empty set. Similarly, some models may not have agent-specific fixed attributes or parameters.}
\begin{equation}
\{{\mathbf{x}_{i,t}}\}_{i=1}^{N}, \mathbf{y}_t \sim \mathbb{P}\left(\{{\mathbf{x}_{i,\tau<t}}\}_{i=1}^{N}, \{\mathbf{y}_{\tau<t}\}, \{\mathbf{a}_i\}_{i=1}^{N}, \boldsymbol{\theta}\right).
\label{eq:abm}
\end{equation}
This definition captures in what sense the model is non-equilibrium, as model variables can be obtained as a map of previously computed variables, rather than by solving equilibrium equations \citep{pangallo2024equations}. However, it characterizes many high-dimensional stochastic processes that are not necessarily ABMs, so one can add additional conditions such as that model units are heterogeneous\footnote{Typically, $\mathbf{x}_{i,t} \neq \mathbf{x}_{j,t}$ for some $i$, $j$ and $t$.} and interact\footnote{Formally, for some $i \neq j$,  $\mathbf{x}_{i,t>0}$ is not independent of $\mathbf{x}_{j,\tau}$ for at least one $\tau \in \left\{0, \dots, t\right\}$.}.

\paragraph{Examples. Why is Schelling's model an ABM, but Brock and Hommes (1998) not?} The \cite{schelling1971dynamic} segregation model, one of the most famous ABM in the social sciences, features $N$ distinct agents, with $N$ typically between tens and thousands. Each agent $i$ has a position $\mathbf{x}_{i,t}$ and a race $\mathbf{a}_i$ that is fixed. At each time step, agents update their position depending on the fraction of neighboring agents with same race at the previous time step. The decision to move depends on a global tolerance threshold $\boldsymbol{\theta}$. This model meets all requirements of our definition. 
In contrast, while the  \cite{brock1998heterogeneous} asset pricing model\footnote{
The \cite{brock1998heterogeneous} paper is a seminal work that models a market where traders exchange a single asset using predefined strategies that adapt over time based on past performance. The model demonstrates that simple rules can lead to complex dynamics in asset pricing.
} is often considered an ABM, it does not meet the definition since it models a continuum of agents and has only two to four trader \textit{types}.

\subsection{What makes an ABM data-driven?}
\begin{figure}[!h]
\centering
\includegraphics[width=0.6\textwidth]{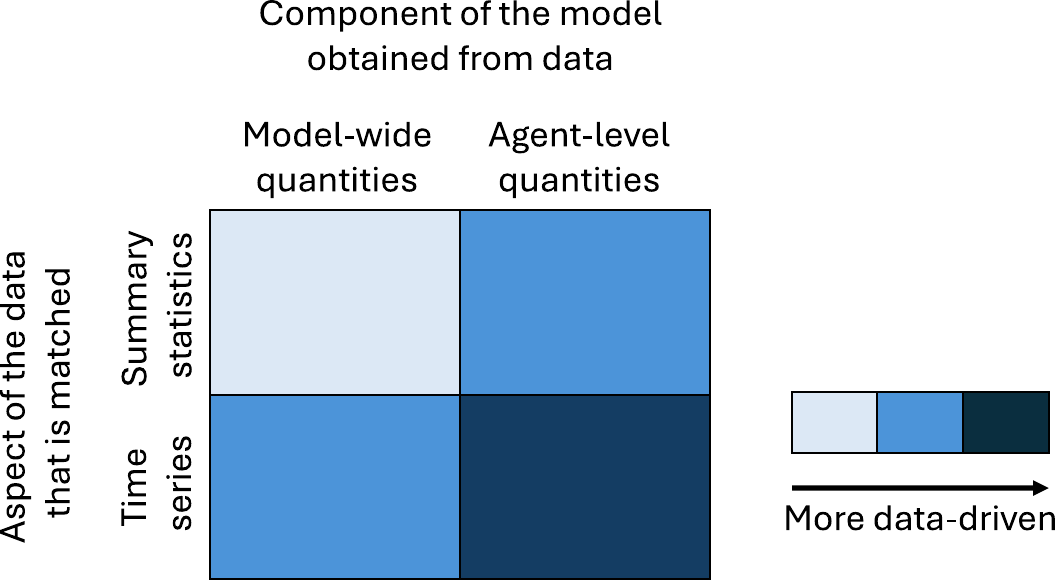}
\caption{Data-driven ABMs classification diagram. We consider two dimensions to measure the extent to which an ABM is data driven. On the vertical axis, whether the model reproduces real world time series or only summary statistics. On the horizontal axis, whether quantities obtained from data are model-wide or agent-specific. We consider all ABMs in the diagram except the ones in the top left quadrant to be data-driven. Darker shades indicate a more data-driven ABM. }
\label{fig:data-driven}
\end{figure}

Aren't all ABMs data-driven? If we consider that all ABMs have been built with an eye to explaining empirical patterns in real-world data, then, yes. For instance, \cite{schelling1971dynamic} built his famous ABM because he wanted to explain racial segregation patterns in US cities. But, does this make the Schelling model data-driven? To address this ambiguity, we think that we need to distinguish between a qualitative and quantitative match with data. We name \textit{theory-driven} ABMs that build theories to qualitatively match patterns in the data.

For ABMs to be data-driven we require that some of the components of Eq. \eqref{eq:abm} are obtained by quantitatively matching the data. The extent to which models are data-driven depends both on what components of Eq. \eqref{eq:abm} are obtained from the data, and on what aspect of the data they aim to match (see Figure \ref{fig:data-driven}). Models that only calibrate parameters are less data-driven than models that also calibrate agent-level attributes and variables. Similarly, models that only match time-independent summary statistics (defined in the next section) are less data-driven than models that track empirical time series, either for in-sample fitting or for out-of-sample forecasting. Models that both obtain agent-specific quantities from data and track empirical time series are the most data-driven (see the color scale in Figure \ref{fig:data-driven}).

Although all models in Figure \ref{fig:data-driven} are to some extent data-driven, to make sense of the developments in methods and applications that occurred in the last 10 years we reserve the ``data-driven'' name only for some models. We say that an ABM is \textit{data-driven} when at least one of the agent-specific variables $\mathbf{x}_{i,t}$ or attributes $\mathbf{a}_{i}$ is obtained from data for all agents $i$, or when the model tracks at least one empirical time series. In other words, we consider all quadrants of Figure \ref{fig:data-driven} except the one on the top left.\footnote{%
In deciding whether a model is data-driven, the unit of analysis is the paper, not the ABM itself. It is in theory possible that an ABM is presented in an abstract way in one paper, and made data-driven in another paper. Although this is possible, we are not aware of any instance. A possible reason is that  setting agent-level quantities from micro data or tracking empirical time series usually requires making specific assumptions while building the ABM that may not be following behavioral theories strictly. By contrast, theory-driven ABMs tend to prefer to follow economic and behavioral theories more closely.
}
\paragraph{Examples. How did macroeconomic data-driven ABMs emerge?} Most macroeconomic ABMs prior to 2015 \citep{dawid2018agent} aimed at replicating stylized facts and summary statistics such as cross-correlations between macroeconomic variables and tail exponents of the firm size distribution. In these models, some parameters were directly taken from data, such as the 2\% target inflation rate that has been used by central banks in the last 30 years. 
However, typically, agent-level quantities were initialized by sampling from theoretical distributions rather than from data. Moreover, the model time series were mostly compared to statistical features of real-world time series, such as matching an average or volatility, rather than attempting to track the series at each time step. While several of these models were fundamental for current data-driven ABMs, the relation to data places them in the top left quadrant of Figure \ref{fig:data-driven}, so they do not fall within the data-driven ABMs classification. 

Data-driven models started to emerge in the last ten years. For instance, the ABM in \cite{papadopoulos2019income} tracks empirical time series even when agent-level quantities are not initialized from data, placing this model in the bottom left quadrant of Figure \ref{fig:data-driven}. Conversely, \cite{otto2017modeling} initialize the variables in their ABM from data, but do not track or forecast real-world time series, placing the model in the top right quadrant of Figure \ref{fig:data-driven}. More recently, \cite{poledna2023economic} initialized their model with agent-level quantities from several real-world datasets, and attempted to track and forecast empirical time series. This model is in the bottom right quadrant of Figure \ref{fig:data-driven}, making it most data-driven. 

\paragraph{Why this definition?}
This definition of data-driven ABMs\footnote{Is ``data-driven ABM'' the right terminology? In many cases, ``data-driven'' indicates statistical methods that do not make almost any assumption about the system being modeled. By contrast, ABMs are theoretical models (also known as mechanistic, or structural, or causal models depending on the discipline) that make lots of assumptions about agents' behavior and interactions. In fact, we like this terminology precisely because of the contrast it creates. Models are data-driven, because data availability dictates many important choices, but they are also agent-based, requiring the modeler to still develop a theory.
} comes from reviewing a large set of economic ABMs and aiming to capture the aspects that made ABMs more likely to resemble the real economy.
While other definitions may be valid, several can be distilled into the two key dimensions we propose.
For instance, one could measure how data-driven an ABM is by the number of data points used, irrespective of whether they are global parameters or agent-specific quantities. Yet, typically there are many more agent-specific quantities than parameters, so initializing at least one quantity for all agents is likely to use more data points than taking all parameters from data. Validation methods could be another criteria, with data driven ABMs being those that can forecast rather than just reproduce  stylized facts. However, a necessary condition for forecasting is to follow time series, and so this criterion reduces to a special case of our classification.
This definition also highlights two important features of ABMs: heterogeneous agents and dynamics. Requiring agent-specific initialization emphasizes heterogeneity, as model-wide quantities alone would suffice if all agents were identical. Focusing on matching time series rather than summary statistics allows to describe transient states in more detail---a feature distinguishing ABMs from equilibrium models.

\section{Methods}
We use different methods depending on which component of Eq. \eqref{eq:abm} we want to obtain from the data. If we care about parameters $\boldsymbol{\theta}$, we do \textit{calibration}. If we want to estimate agent-level attributes $\mathbf{a}_{i}$ and initial conditions $\mathbf{x}_{i,0}$, we do \textit{initialization}. If we want to estimate the entire time series of agent-specific variables, $\mathbf{x}_{i,t}$, we do \textit{data assimilation}.

\subsection{Parameter calibration}
The oldest and most developed method is parameter calibration.\footnote{
Sometimes we talk about parameter estimation. In principle, estimation is about finding the correct values of parameters, while calibration is about finding a value that makes the model match the data (whether it is the correct value or not), but the two terms are often used interchangeably. Here, to avoid confusion we just refer to calibration.
} Parameter calibration was developed for models in the top left quadrant of Figure \ref{fig:data-driven}, which we do not strictly consider data-driven, but calibration methods are also useful for parameters in more data-driven models. 

When possible, parameters are directly derived from data. For instance, suppose that one of the parameters of the model is the percentage reduction in the asking price of a flat after it has not been sold for two months. We may look at housing market data or at the literature and pick a value, say 5\%. Alternatively, we can rely on previous estimations or expert knowledge. For example, \cite{pichler2022forecasting} used the parameters suggested in \cite{muellbauer2020} out of educated guesses to set changes in consumption patterns during the Covid-19 pandemic. For parameters where there is no measurement or reliable expert knowledge, one can infer them from data in four steps: (i) sampling, (ii) choice of summary statistics, (iii) definition of the loss and (iv) criterion to select the parameters. Most calibration methods are a combination of choices at these four steps.

\paragraph{Sampling.}  
A basic approach to sampling the parameter space is uniform random sampling within reasonable bounds. While this may suffice for smaller spaces, larger parameter spaces often require more sophisticated methods to ensure adequate coverage. Techniques like Nearly Orthogonal Latin Hypercubes and low-discrepancy sequences, such as Halton sequences \citep{borgonovo2022sensitivity}, help by spreading the sampling points more evenly across the space, reducing gaps and clustering. However, when model evaluations are computationally expensive, these methods, which set sample before evaluating, may also become inefficient or impractical. In such cases, adaptive methods offer a more efficient alternative by iteratively selecting promising points based on previous evaluations. Examples include the Nelder-Mead simplex search algorithm \citep{franke2009applying}, as well as newer techniques that employ \textit{meta-models} (also known as surrogate models or emulators) to predict performance at unsampled points \citep{lamperti2018agent,glielmo2023reinforcement}.

\paragraph{Summary statistics.} We define a summary statistic as any scalar or vector that reduces the dimensionality of simulated and real data across time to make them easier to compare. A popular method that uses summary statistics is the simulated method of moments \citep{delligatti2018agent}. Originally, this method consisted of calculating longitudinal statistical moments of economic variables, such as mean, variance, and skewness of unemployment. However, the terminology has been extended to any other scalar quantity that can be computed in both real and simulated data. For instance, a moment could also be the total number of firms that default in the considered period. However, one does need to restrict to statistical moments when studying steady states or pseudo steady states, other choices include parameters of an auxiliary model (indirect inference) or the distribution of an important variable when the model reaches a pseudo steady state \citep{kukacka2017estimation}, where variable distributions are stationary. One can be even more creative, and consider information-theoretic measures that capture up-down patterns in time series \citep{barde2017practical} or similarities in causal structures \citep{guerini2017method}.

\paragraph{Definition of the loss.} Once we have computed summary statistics in simulated and real data, we must combine them to obtain a total loss. This can be done in several ways, the most principled way is to use the method of simulated moments and weigh each moment by the inverse of its uncertainty (as captured by the covariance matrix), but equal weighting of all summary statistics is also common. When the summary statistic is the distribution at the pseudo steady state, the loss is the simulated likelihood \citep{kukacka2017estimation}.

\paragraph{Criterion to select the parameters.} Once we have a total loss associated to each parameter combination, we may take the frequentist approach and pick the parameter combination that minimizes the loss, using any optimization algorithm, including gradient-based, gradient-free, simulated annealing and genetic algorithms. Alternatively, we could take a Bayesian approach and select parameter combinations with a probability that is inversely proportional to the loss. Common Bayesian methods include Approximate Bayesian Computation \citep{pangallo2023unequal}, Kernel Density Estimation \citep{grazzini2017bayesian} and Neural Posterior Estimation \citep{dyer2024black}.

\paragraph{So, what calibration method should one choose?} With several calibration methods available, it can be difficult to pick one. Unfortunately, there are very few systematic comparisons of which methods work best in which setting, and these comparisons are mostly inconclusive. For instance, \cite{platt2020comparison} shows that Bayesian methods work better than frequentist methods in four simple models, but \cite{carrella2021no}, considering 41 models, shows that no method clearly outperforms the others. To navigate the wilderness of calibration methods, our suggestion is twofold. First, we think that the choice of summary statistics, often overlooked, is crucial, as it strongly influences which parameters can be identified.\footnote{For instance, in a labor market ABM, if you only use aggregate unemployment rates and wage distributions, you might miss parameters that govern job search intensity or firm-specific wage setting. By including job search duration, vacancy fill rates, and wage negotiation outcomes in the summary statistics, you can better identify both worker search strategies and firm hiring behavior.}. Second, we suggest to choose the method that minimizes the combination of human and computer time that works best for the modeler, taking advantage of software packages such as \texttt{Black-IT} \citep{benedetti2022black} or \texttt{BlackBIRDS} \citep{quera2023blackbirds}. 

\subsection{Initialization}
\label{sec:initialization}
When we use the term initialization we refer to agent-level quantities, namely attributes $\mathbf{a}_{i}$ and initial conditions for variables, $\mathbf{x}_{i,0}$.\footnote{
The reason why we include attributes, which are fixed in time, in the scope of initialization, which generally refers to time-varying quantities, is that the techniques used to determine $\mathbf{a}_{i}$ and $\mathbf{x}_{i,0}$ are similar.
} Until few years ago, the typical approach in economic ABMs was to initialize variables and attributes at random. The last few years saw more attempts to initialize them from data, with a mix of principled and unprincipled methods. Initialization requires to make disparate sources of data compatible with themselves and with the model. This task requires lots of modeler attention, deep knowledge of economic statistics and national accounting, and many lines of code. In a data-driven ABM project, it is not unusual to develop more code on initialization than on the ABM itself. In the following, we give some examples of initialization procedures.

\paragraph{Generating synthetic populations.} Data-driven ABMs often focus on a specific population of households, firms, banks, etc. In most cases, the population belongs to a geographical area, but it could also correspond, for instance, to an industry (e.g. pharmaceutical firms). In either case, the synthetic population of agents must match the real population. The main difficulty is to respect the joint distribution of agents' attributes and initial conditions. Luckily, this is not a new problem, as it has already been extensively investigated in several social sciences including epidemiology, land use, urban science and transportation research \citep{arentze2007creating}, and in economics by the microsimulation community \citep{li2013survey,richiardi2024back}. Principled methods, such as Iterative Proportional Fitting and Combinatorial Optimization, are often used in conjunction with ad-hoc, case-specific methods. 

\paragraph{Network reconstruction.} Attributes and initial conditions may also include the network connecting the agents. Here, most efforts in economics have focused on reconstructing production \citep{ialongo2022reconstructing,mungo2023reconstructing} and financial \citep{anand2018missing} networks. There is also a large literature on regionalization of input-output models \citep{bonfiglio2008assessing}.

\paragraph{Compatibility with national accounts.} National accounts are a great source of data to initialize data-driven ABMs, especially when ``agents'' are industries \citep{pichler2022forecasting}. They also serve as a benchmark that agent-level data must be consistent with---for instance, the initial conditions for agent-level consumption must sum to total consumption. The problem is that agent-level data are often available as surveys, may not consider the same sample as national accounts, and may use a different classification system and aggregation level. Sticking to consumption as an example, \cite{pangallo2023unequal} used data from a US consumption survey that were available for detailed income and age subgroups, but were largely inconsistent with national accounts. For instance, imputed rents count as consumption, but the survey did not ask for this spending item. \cite{pangallo2023unequal} used bridge tables and ad-hoc assumptions to make the data compatible with national accounts, but there exist more systematic attempts \citep{cazcarro2022linking}. 

\paragraph{So, what should I do regarding initialization?} Roll up your sleeves, search for data, investigate in detail how the data are constructed, use principled methods when available, and your intuition for which approximations work best when better data and methods are not available. For instance, if data on consumption basket by household income are not available for a country, it may be a good strategy to use the data from another country that is similar. In our experience, approximations of this kind only change initialization of some agent-level variables by a few percentage points, barely affecting aggregate results.

\subsection{Data assimilation}
\label{sec:dataassimilation}

Data assimilation is about estimating the entire time series of agent-specific or global variables, $\mathbf{x}_{i,t}$ and $\mathbf{y}_{t}$ that are not observed, i.e. \textit{latent} variables.\footnote{While in deterministic models one only needs to estimate the initial condition $\mathbf{x}_{i,0}$ and $\mathbf{y}_{0}$, in ABMs, which are typically stochastic, this is not sufficient. Thus, data assimilation estimates the entire sequence of latent agent-specific and global variables.} This is particularly important when studying the timing of effects is of essence.  For example, consider a macroeconomic ABM that focuses on business cycles. A well calibrated ABM may be able to reproduce the frequency of recessions. However, to match the exact timing of recessions, we must know the value of relevant latent variables. For instance, to match the timing of the 2008 recession we must correctly estimate over time latent variables such as imbalances in the financial system. Conversely, data assimilation may not be important when tracking empirical time series is not crucial, for instance when focusing on scenarios in the future. In the following, we review the main data assimilation methods.

\paragraph{Kalman and particle filters.} Rooted in Bayesian statistics , Kalman and particle filters are used to combine models and observations \citep{carrassi2018data}. These filters adjust the latent variables so that the model produces observations close to the real world. The ensemble Kalman filter, a specific type of Kalman filter that only requires to simulate the model with no need to compute its Jacobian (as opposed to the standard Kalman filter), was introduced in ABMs by \cite{ward2016dynamic}, and recently applied to inequality dynamics \citep{oswald2024agent}. The particle filter, which is more flexible but also more computationally expensive than the ensemble Kalman filter, has been used in economics with heterogeneous agents financial market ABMs \citep{lux2018estimation}. 

\paragraph{Probabilistic graphical models.} Ensemble Kalman and particle filters treat the ABM as a black-box. However, the dependency structure of model variables carries useful information. If it is possible to represent the ABM as a probabilistic graphical model, one can take advantage of conditional independence relations between variables to write a computationally tractable likelihood function of the latent variables. \cite{monti2020learning} applied this formalism first, to an opinion dynamics ABM, while \cite{monti2023learning} apply it to an economic ABM of the housing market.

\paragraph{Heuristic methods.} It is common to initialize the latent variables of the model (that cannot be initialized using the techniques in Section \ref{sec:initialization}) by running the model for a transient period until it settles to a quasi steady state, and then discard the transient period. In this way, the latent variables should become compatible with the observed variables \citep{geanakoplos2012getting}. Moreover, when part of the aggregate variables $\mathbf{y}_{t}$ are observed, they can be taken directly from the data, so the model is fit with an exogenous trend \citep{papadopoulos2019income}. This is particularly useful in case of abrupt disruptive events, such as natural disasters or pandemics. For instance, \cite{pichler2022forecasting} impose exogenous trends on the Covid-related restrictions faced by different industries, matching the restrictions that were actually placed by the UK government in the spring of 2020.

\paragraph{So, what data assimilation method is best?} It is often a good idea to use exogenous trends, when they are available. But certain trends may be latent, for instance opinions or beliefs. In this case, the only option is to use one of the methods listed above. The literature on data assimilation in ABMs is not yet mature enough for general guidelines, but our conjecture is that filters work best when only aggregate data are observed, while the representation of an ABM as a probabilistic graphical model is particularly useful when some agent-level variables are observed.

\section{Success stories}
We now discuss applications, highlighting the value added of making ABMs data-driven in obtaining domain-specific insights. 
Examples throughout are chosen to illustrate success stories of data-driven ABMs, rather than to provide an exhaustive review.

\paragraph{Housing markets.}  
To understand what could have stopped the housing market bubble that led to the 2008 crisis, 
\cite{geanakoplos2012getting} built the first data-driven ABM of the housing market.\footnote{See \cite{axtell2014agent} for a more detailed description of the model.} The key result is that raising interest rates would have done little to prevent the bubble, while imposing stricter limits on loan to value would have been much more successful. What makes this model's counterfactuals trustworthy is that the model was able to reproduce the dynamics of the Washington, D.C. housing market from 1997 to 2009 with actual central bank implemented policies. Several central banks have now used and extended this model, see for instance \cite{mero2023high} and \cite{borsos2024agent} in this volume. 

More recently, spatially explicit data-driven models have been developed to assess how physical risk, often driven by climate change, affects the housing market \citep{filatova2015empirical,pangallo2024climate}. For instance, \cite{haer2017integrating} find that under bounded rationality, households in the Netherlands will adopt mitigation measures that halve expected annual damage from floods compared to a no mitigation scenario. The quantitative reliability of this result hinges on using data from a specific area with detailed projections of how flood risk evolves under climate change. 

\paragraph{Labor markets.} 
Data-driven ABMs have modeled labor frictions using empirical data and network structures, and studied how net-zero policies and emerging technologies reshape labor markets. For instance, \cite{axtell2019frictional} show that using empirical labor flow networks, which are heavy tailed, raises unemployment and shifts the Beveridge curve \footnote{The Beveridge curve is an empirical relationship between job vacancies and unemployment, typically depicted as a negatively sloped curve. Shifts in the curve’s position can signal changes in labor market efficiency, frictions, or other structural factors influencing the matching of workers to jobs.} toward higher unemployment for the same vacancy rate. At the occupational level, \cite{del2021occupational} demonstrate that while some occupations face higher automation risk, others with fewer transition paths may actually experience greater unemployment. This model also shows that moving beyond equilibrium assumptions reproduces the Beveridge curve’s counterclockwise cyclical behavior during business cycles. Building on these approaches, \cite{fair2023emerging} incorporate industry and geographic mobility frictions into a data-driven ABM, and \cite{berryman2024skill} studied how geographic constraints shape growth and sustainable development scenarios in Brazil. See chapter \cite{del2024enhancing}for complementary discussions on agent-based models and labor markets.

\paragraph{Economic impact of natural disasters and pandemics.} Quantifying the impact of disasters is essential for policy making but challenging. Models require estimating both direct disruptions as well as indirect propagation of shocks across the production network. By leveraging industry input-output or firm-level supply chain data, these models quantify the relative importance of indirect effects compared to direct effects.
\cite{hallegatte2008adaptive}'s pioneering work estimates the impact of hurricane Katrina on the economy of Louisiana. By initializing the input-output network connecting industries in the model with actual data, the author finds that the indirect impact resulting from the propagation of direct shocks amounts to 50\% of the direct impact, a substantial propagation. This model was able to match the recovery profile of employment and industries for a year following the hurricane. Subsequent models have broadened this approach to incorporate international trade \cite{otto2017modeling}, firm-level heterogeneity \cite{inoue2019firm}, diverse households \cite{markhvida2020quantification}, and transportation networks \cite{colon2021criticality}.

A similar modeling framework has been used to understand the economic impact of the Covid-19 pandemic. Thanks to careful initialization on real-world data, \cite{pichler2022forecasting} made a forecast on the UK economy recession in the second quarter of 2020 that turned out to be more accurate than those made by most public and private institutions. This showcased for the first time how data-driven ABMs could be used for forecasting ahead of time (that is, the forecast was made before data on the performance of the UK economy were released). \cite{pangallo2023unequal} extended this model and merged it with a data-driven epidemiological ABM. The resulting model simulates the epidemiological and economic decisions of a synthetic population of half a million people in the New York metropolitan area. The model accurately fits the dynamics of the first lockdown in spring 2020, reproducing a stylized fact typical of the Covid-19 recession that more traditional models had more difficulty replicating, namely that high-income people got less unemployed but reduced consumption more than low-income people. In addition, the model uses counterfactuals to show that both epidemiological and economic impacts are similar whether policy makers impose lockdowns or the population spontaneously changes their behavior out of fear of infection, reducing contacts and consumption. Thus, under behavior change, it is the virus, not the lockdown, that takes a toll on the economy.

\paragraph{Macroeconomics.} The first data-driven macroeconomic ABM that we are aware of is \cite{bergmann1974microsimulation}.\footnote{The author uses the term ``microsimulation'', but the model could be described as an ABM ante litteram \citep{richiardi2024back}.} This is a fully fledged macroeconomic ABM, with workers/consumers, firms, a bank, a government and a central bank. The firms, representing industries, are connected through an input-output network initialized from real data. The author shows that the model is capable of fitting U.S. macroeconomic statistics from the beginning of 1967 to the end of 1970. This model was  visionary and ahead of its time. Around 30 years later, agent-based macroeconomics started to develop, obtaining important theoretical results on growth and business cycles but without connecting much to real-world data \citep{dawid2018agent}. This trend changed in recent years \citep{papadopoulos2019income, kaszowska2020macroprudential}. The model that so far had most impact is \cite{poledna2023economic}, who represented the Austrian economy at a one-to-one scale and showed that the ABM could compete with statistical vector autoregressive (VAR) models and traditional dynamic stochastic general equilibrium (DSGE) models at out of sample forecasting. This model has now been used for multiple applications. For instance, it showed that post-pandemic inflation in Canada was initially due to increases in input costs, then to expectations of future inflation, and finally by increases in demand by other firms as the economy reopened after lockdowns \citep{hommes2024agent}.

\section{Challenges and Opportunities -- Outlook and ways forward}

How does the evolution towards data-driven ABMs affect the way we validate models, choose behavioral rules and build counterfactuals? We discuss answers to these questions and challenges and opportunities lying ahead.

\subsection{Data, methods and validation}

\paragraph{Developing new methods.} We have covered several calibration, initialization and data assimilation methods that are now available for modelers. While calibration techniques have been developed for more than a decade, the initialization and data assimilation methods are fairly new for ABMs.  A key challenge is enhancing these methods to adjust all micro-variables in large-scale models, which is essential for macroeconomic ABMs to accurately track long-term time series. There is a growing literature exploring alternatives. For example, \cite{grattarola2021learning,casert2024learning,cozzi2024neural}  use neural network architectures as surrogate models when the ABMs are too complicated to estimate their latent variables. Developing these methods further presents significant opportunities to improve the performance and applicability of ABMs.

\paragraph{Dealing with massive datasets is time-consuming.} A second, related, challenge is that initialization is time consuming, since it requires finding reliable data sources and then making them compatible with themselves and with the model. One way forward relies on governments and statistical offices facilitating access to data, as well as possible collaboration with private sector for additional data (see \citealt{turrell2024cutting} for a more detailed discussion). Another suggestion, which has worked in our experience, is to work with large teams, where for instance two or three PhD students work on different aspects of the same project. This transformation towards a science-based lab model may already be happening in empirical economics \citep{athey2018impact}, and could be effective for data-driven ABMs as well.

\paragraph{Pushing validation to tracking time series.} The main opportunity from using advanced methods and detailed data is to push forward the concept of validation. Traditionally, validation of ABMs has focused on replicating stylized facts and broad empirical regularities. Of course, this comes at the risk of overfitting, as complicated models with many parameters can easily fit a few stylized facts. Recently, more rigorous validation has been possible through out-of-sample forecasting. In this approach, modelers calibrate the model on some initial year(s) and then validate it on its ability to forecast economic dynamics in future years that have not been considered in the calibration process. Although computationally and data demanding, several ABMs have succeeded in doing this. For example,  \cite{pichler2022forecasting} used their ABM to forecast the UK’s industry-level output during the COVID-19 pandemic ahead of time. \cite{poledna2023economic} demonstrated that their ABM competed with Vector Autoregressive (VAR) and Dynamic Stochastic General Equilibrium (DSGE) models in predicting Austria’s GDP. Going forward, we consider that stylized facts should be the minimum bar for ABMs. When possible ABMs should try to seek validation through out of sample forecasting, or at least in sample fitting time series. This validation benefits from fine grained data and advanced statistical methods and hence developing these methods and acquiring data should a priority for the field.

\subsection{Data, behavior and theory}

\paragraph{A general boundedly-rational approach for modeling behavior.} Modeling behavior is hard. The main advantage of the optimization framework traditionally used in economics is that it is a one-size-fits-all approach---one can use optimization to model decisions of consumers, firms, banks, politicians, even parents when making family choices. We currently lack an alternative general bounded rationality framework \citep{harstad2013bounded}. This limitation led to criticism that ABMs rely on too many arbitrary ``ad-hoc'' assumptions and free parameters.\footnote{Arbitrary choices also exist in models using optimization, they are just hidden in the functions that are optimized. For example, DSGE models sometimes decide between preference structures, such as King–Plosser–Rebelo (KPR) or Greenwood–Hercowitz–Huffman (GHH) preferences, depending on practical modeling needs rather than realism. Similarly, when is it reasonable to use ``MIT-shocks''? How about ``iceberg costs''? } 
Recent efforts have explored alternatives based on  laboratory experiments \citep{hommes2021behavioral}, psychology \citep{roos2018values}, economic theory \citep{gabaix2019behavioral}, and Large Language Models \citep{horton2023large,del2024llms}. These approaches are promising and further development of these methods could lead to one or a few standard frameworks in the future.

\paragraph{Data can replace unnecessary assumptions.} Data-driven ABMs can address the ``ad-hoc'' critique in two ways. First, they replace assumptions that are not critical to the model with real-world data. For example, in studying policies to reduce the economic impact of COVID-19, it is not essential to model detailed individual choices, such as how agents decide on commuting patterns or places to go out. Instead, one can use mobile phone data to infer people's mobility and contacts with others, and model the reduction in contacts due to the fear of infection with just one parameter \citep{pangallo2023unequal}. This results in fewer choices and free parameters.\footnote{This approach could be criticized by agent-based modelers who think that all results should be ``grown'' from first principles \citep{epstein1996growing}. We agree that at least some assumptions should come from theory, but we also think that it is a fair compromise to draw less important assumptions from data.}

\paragraph{The contribution of assumptions to model validity can be quantitatively tested.} A second opportunity for data-driven ABMs to address the ``ad-hocness'' critique is through measuring which assumptions really improve results, and drop unnecessarily complicated assumptions when they do not. Agent-based modelers are often proud of making ``reasonable'' assumptions. \cite{farmer2024making} calls this the principle of \textit{verisimilitude}: ``assumptions should be plausible. Assumptions that seem wrong from the outset are more likely to lead to false conclusions than plausible assumptions''. But who decides which assumptions are plausible and which assumptions are not? Inspired by machine learning, in the last few years psychologists have been making comparisons based on forecasting \citep{erev2017anomalies, artinger2020taxi}. Data-driven agent-based modelers should follow in their footsteps, systematically testing how their assumptions may improve on the validity of the model, for instance its ability to do out-of-sample forecasting.

\subsection{Data, external validity and counterfactuals}

\paragraph{External validity should be addressed.} One potential drawback of heavily relying on data is the criticism of external validity, ``how do you know that insights from your model also hold beyond the region/country/industry you have modeled?''. This is a fair criticism shared with all empirical work. To understand the extent to which a data-driven ABM can yield insights on other regions/countries/industries of interest, the modeler should study how data change between systems of interests. For example, input-output coefficients vary little between regions of the same country, but it may be useful to check that results do not critically depend on the specific values of input-output coefficients.

\paragraph{Counterfactuals are more reliable.} The problem of external validity highlights a substantial opportunity for data-driven ABMs, compared to purely abstract ABM that does not reproduce any specific system. By reproducing the dynamics of, say, a region, under the policy interventions that were in place, studying counterfactual policies yields much more reliable results \citep{geanakoplos2012getting, pangallo2023unequal}. Data-driven ABMs should always aim at reproducing the actual dynamics as a baseline, and then test how alternative policies may have led to different outcomes in some specific historical episode.

\section{Conclusion}
The main takeaways of the paper are that data-driven ABMs: (i) push validation standards from stylized facts to time-series tracking and forecasting (ii) help model behavior in a general way by replacing unimportant assumptions with data and making it possible to systematically test the impact of behavioral assumptions on model outputs; (iii) make counterfactuals more reliable by reproducing empirical dynamics with actual policies. These advantages address current limitations of agent-based models, and will promote their use in economic research and policy. Accurate models can inform policies that reduce unemployment, control inflation, and improve overall well-being.

Going forward, we must bear in mind that data-driven ABMs will not make economics value-free. While validation may protect models from arbitrary assumptions and increase objectivity, complete objectivity in economics is an illusion \citep{atkinson2009economics,coyle2021cogs}. Economic decisions, models, and policies reflect underlying ethical and normative judgments, and economists should openly acknowledge these values and engage in transparent discussions about them. Incorporating insights from other disciplines, such as sociology, ethics, and political science, can further help address the ethical and societal implications of economic analysis.

\bibliographystyle{econ}
\bibliography{bibliografia}

\end{document}